\documentclass[lettersize,journal]{IEEEtran}
\usepackage{amsmath,amssymb,amsfonts}
\usepackage{algorithmic}
\usepackage{algorithm}
\usepackage{array}
\usepackage[caption=false,font=normalsize,labelfont=sf,textfont=sf]{subfig} 
\usepackage{textcomp}
\usepackage{stfloats}
\usepackage{url}
\usepackage{verbatim}
\usepackage{graphicx}
\usepackage{cite}
\usepackage[table]{xcolor} 
\usepackage{epstopdf}
\usepackage{balance} 
\usepackage{caption}
\def\BibTeX{{\rm B\kern-.05em{\sc i\kern-.025em b}\kern-.08em
    T\kern-.1667em\lower.7ex\hbox{E}\kern-.125emX}}

\begin{document}

\title{Fluid Volume Assignment for Flow-Based Biochips: State-of-the-Art and Research Challenges}

\author{Alexander Schneider, Jan Madsen and Paul Pop
\thanks{P. Pop, and J. Madsen are with DTU Compute, Technical University of Denmark, 2800 Kongens Lyngby, Denmark (e-mail: paupo@dtu.dk). A. Schneider was with the same group when this work was done.}%
}

\markboth{Preprint submitted to arXiv}{Schneider \MakeLowercase{\textit{et al.}}: Fluid Volume Assignment for Flow-Based Biochips}

\maketitle


\begin{abstract}
Microfluidic biochips are replacing the conventional
biochemical analysers integrating the necessary functions on-chip. We are interested in Flow-Based Microfluidic Biochips (FBMB), where a continuous flow of liquid is manipulated using
integrated microvalves. Using microvalves and channels, more complex Fluidic Units (FUs) such as switches, micropumps, mixers and separators can be constructed. When running a biochemical application on a FBMB, fluid volumes are dispensed from input reservoirs and used by the FUs. Given a biochemical application and a biochip, one of the key problems which we are discussing in this paper, is in determining the fluid volume assignment for each operation of the application, such that the FUs' volume requirements are satisfied, while over- and underflow are avoided and the total volume of fluid used is minimized. We illustrate the main problems using examples, and provide a review of related work on volume management. We present algorithms for optimizing fluid volume assignments and for reusing leftover fluids to reduce waste. This also includes the optimization of mixing operations which significantly impact the required fluid volumes. We identify the main challenges related to volume management and discuss possible solutions. Finally we compare the outcome of volume management using fixed- and arbitrary-ratio mixing technology, demonstrating significant reductions in fluid consumption for real biochemical assays.
\end{abstract}


\section{Introduction}

Microfluidics refers to a technology that miniaturizes biological and chemical processes to a sub-millimetre scale. A biochip integrates different biochemical functionalities such as mixers, incubators, filters and detectors on a single chip, leading to higher portability, throughput, sensitivity and reduced fluid requirements. Microfluidic Very Large-Scale Integration (mVLSI) enables the development of microfluidic chips using hundreds of such functions, allowing multiple assays to be run in parallel, making them usable for tasks such as Protein Crystallography, Amino Acid Analysis or Chemical Synthesis~\cite{daniel10}~\cite{quake_review_07}. In this paper we are interested in Flow-Based Microfluidic Biochips (FBMB), which manipulate the fluids as a continuous flow. The key for complex functionality in FBMBs is a microvalve~\cite{thorsen-quake02}, similar to transistors in semiconductor VLSI. Such valves are manufactured using multilayer soft lithography and are controlled by outside pressure sources~\cite{minhass:controlsynth}. Using micro valves and channels, more complex Fluidic Units (FUs) such as switches, micro-pumps, mixers and separators can be constructed. Using these valves the flow of fluid within the chip can be restricted, allowing to decide if and in which order the functions on the chips are used. With adequate optimization strategies, a large number of such components can be controlled with only few pressure sources~\cite{Schneider:pincount}.

Ongoing research enables the fabrication of increasingly complex biochips, with an integration density advancing faster than Moore's Law~\cite{hong03}. For example, a commercial LoC featuring 25,000 integrated microvalves that can run 9,216 polymerase chain reactions in parallel has been available since 2008~\cite{perkel:microfluidics}. Many biochip enabled systems are available from companies such as Fluidigm, where even fully automated components can be obtained~\cite{fluidigm}.

The trend is towards programmable multipurpose biochip platforms~\cite{quake_review_07}, which reduce the design, fabrication and testing times, compared to biochips that are custom-built to implement specific protocols and are not programmable. The control signals needed to run an assay are automatically determined by a ``compiler'', which is a software program that converts the source code of the biochemical application (e.g. writen in a language such as Aqua or Biostream) into signals interpretable by control hardware.

Although there has been a lot of work on the design and programming of biochips~\cite{pop2016microfluidic,huang2021computer}, researchers have initially ignored the problem of fluid management~\cite{alistar15}, assuming that all the fluidic constraints are satisfied during the execution of an application. In this paper we illustrate the main problems related to fluid management~\cite{alsch_phd}, including mixing, and review the state-of-the-art in this area. We present novel solutions for fluid volume assignment that minimize reagent consumption through efficient mixing tree generation and leftover fluid reuse, demonstrating significant reductions in fluid usage for real biochemical assays.

The remainder of this paper is organized as follows: In Section~\ref{sect:ArchModel}, we present the biochip architecture model and application model. Section~\ref{sec:mixing} discusses mixing optimization techniques for flow-based biochips. In Section~\ref{Sect:VM}, we present our fluid volume assignment algorithm and demonstrate its application with examples. Section~\ref{sect:experiments} provides an experimental evaluation of our approach on real-life biochemical assays. Section~\ref{Sect:related-work} reviews related work, and Section~\ref{sect:conclusion} concludes the paper.

\section{Biochip Architecture Model}

\begin{figure}[b]
    \centering
        \includegraphics[width=0.4\textwidth]{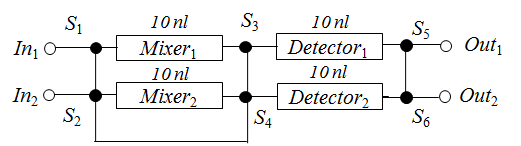}
        \caption{Architecture netlist including two mixers and detectors with a maximum hardware capacity (MHC) of 10 nl each.}
        \label{fig:arch}
\end{figure}

\begin{figure}[h]
    \centering
        \includegraphics[width=0.3\textwidth]{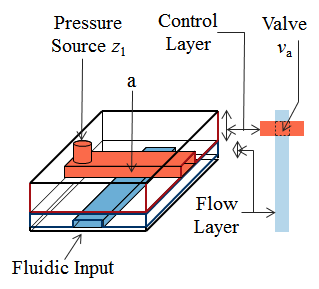}
        \caption{Micro-mechanical valve}
        \label{fig:mecvalve}
\end{figure}

A functional view of a flow-based biochip is shown in Fig.~\ref{fig:arch} which contains multiple \textbf{Functional Fluid Units (FFU)} such as inputs, mixers and detectors, as well as switches connecting the channels leading to these components. The basic building block of these components is a microvalve, which can be used to manipulate the fluid flow.
Fig.~\ref{fig:mecvalve} shows a micromechanical valve which can be manipulated by an external force to either restrict or permit the fluid flow. To do so, the chip is logically divided into two layers, the flow-layer (colored in blue in Fig.~\ref{fig:mecvalve}) which contains the fluid and the control-layer (colored in red) which can manipulate the flow-layer. Such valves are fabricated using a layer of PDMS for each of the two logical layers. The PDMS is poured onto the control- and flow-mold respectively and is then baked. Afterwards the layers are peeled off, aligned and bonded together by further baking, see~\cite{duffy:valve}~\cite{quake:valve} for more details on the fabrication process.

The pressure source ($z_1$ in Fig.~\ref{fig:mecvalve}) is connected to the control channel via a control pin. A control pin is a physical hole, typically significantly larger in diameter than the control channel it is connected to, located at the edges of a biochip. When pressure is applied, the elastic control layer will ``pinch'' the flow layer (at point $a$ in Fig.~\ref{fig:mecvalve}), blocking the flow of fluid (closed valve). If no pressure is applied, the fluid can flow freely through the flow layer (open valve). Hence, this is called a ``normally-open valve''. Normally-closed valves can be fabricated as well and information about their functionality and fabrication is available in~\cite{grover09}; our work can be used with any microvalve technology.

\begin{figure}[h]
    \centering
    \subfloat[Switches]{
    \includegraphics[width=0.3\textwidth]{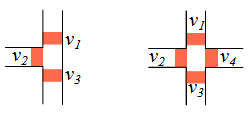}
    \label{fig:switches}
    }
    \newline
    \subfloat[Mixer]{
     \includegraphics[width=0.3\textwidth]{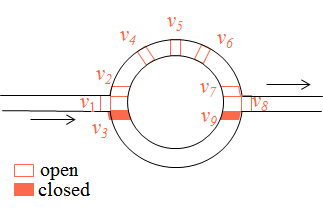}
    \label{fig:mixer}
    }
    \caption{Examples of Fluidic Functional Units}
    \label{fig:microcomps}
\end{figure}

To create functional components, multiple valves are needed. Fig.~\ref{fig:switches} shows two variations of switches, requiring three and four valves respectively to control the path of the fluid entering from any side. Mixers as shown in Fig.~\ref{fig:mixer} require nine valves to be operational. $v_2$ and  $v_7$ as well as $v_3$ and $v_9$ are used to close of one half of the mixer to allow the other half to be filled. Valves $v_1$ and $v_8$ close off the mixer during the mixing process, which is indicated by valves $v_4$, $v_5$ and $v_6$, acting as a pneumatic pump. Other components such as filters, heaters or detectors only require two valves to close off the component during its execution, similar to valve 1 and 8 in the mixer~\cite{chou01}.

\begin{figure*}[h]
    \centering
    \subfloat[Step 1]{
    \includegraphics[width=0.25\textwidth]{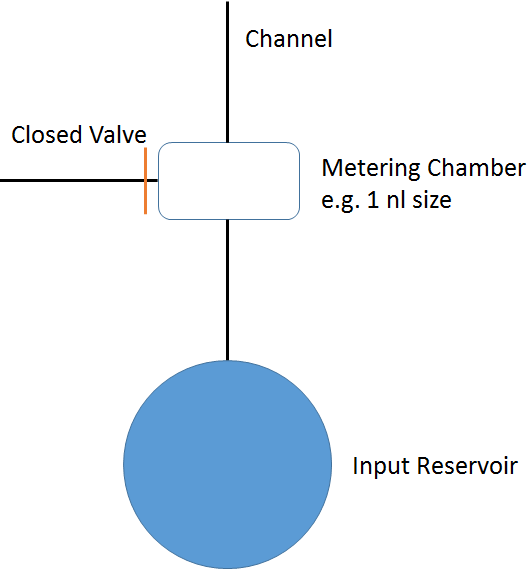}
    \label{fig:meter1}
    }
    \subfloat[Step 2]{
     \includegraphics[width=0.25\textwidth]{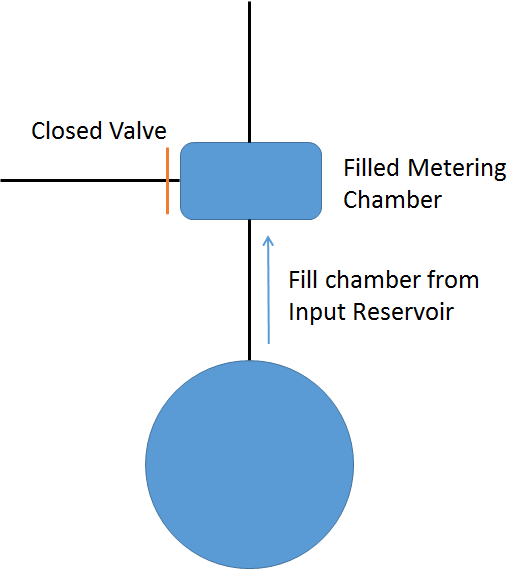}
    \label{fig:meter2}
    }
    \subfloat[Step 3]{
     \includegraphics[width=0.25\textwidth]{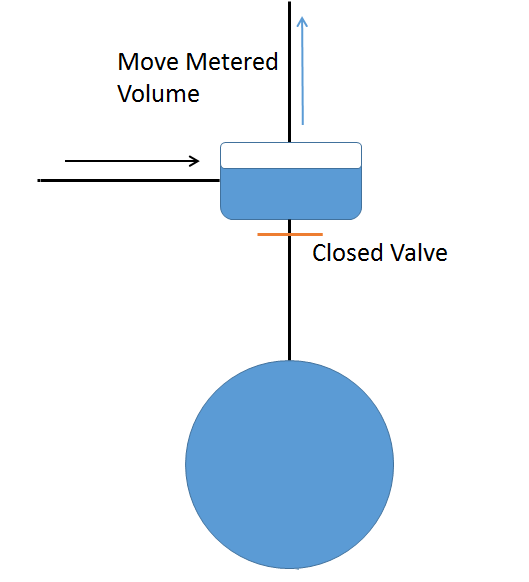}
    \label{fig:meter3}
    }
    \caption{Metering example using a metering chamber}
    \label{fig:Metering}
\end{figure*}

\subsection{Architecture Model}
\label{sect:ArchModel}
For our examples we use the system-level architecture model based on a topology graph, as proposed in~\cite{pop2016microfluidic}. Fig.~\ref{fig:arch} shows an example of a biochip architecture, containing two inputs, outputs, mixers and detectors. We distinguish between two kinds of vertices in this model: Switches, which create intersections between multiple channels (e.g. S$_1$) and FFUs which can perform a certain action (e.g. In$_1$, Mixer$_1$, Detector$_1$). Edges represent channels through which fluid can be transported and are considered bi-directional throughout this paper. Fluid can be transported through these components by applying pressure to an input from an outside source, or by generating force on-chip, e.g. by using a pneumatic pump~\cite{chou01}. For simplicity reasons we do not model pumps in our architectures, but instead assume that an input is capable of providing pressure, as well as being used as a source of reagents. Our proposed method is however not limited to an architecture as such.

Each FFU has a \textbf{Maximum Hardware Capacity (MHC)} specified, which is the maximum volume of fluid that
can be handled by the FFU. These are depicted in nanoliters (nl) on top of each FFU in Fig.~\ref{fig:arch}. The \textbf{Minimum Volume Requirement (MVR)} is the minimal amount of fluid a FFU requires to function properly (e.g. a detector requires a certain amount of fluid for reliable optical detection). The smallest volumes of fluid that can be transported in a given FBMB architecture is the \textbf{Hardware Transport Resolution (HTR)}. Depending on the architecture, the HTR can for example be the smallest amount of liquid that can be correctly metered, or a minimal volume required to fill the depth and width of a channel. Fig.~\ref{fig:Metering} shows a simplified example of a metering chamber, used to obtain a certain volume of fluid from the input reservoir. More information on metering chambers can be found in~\cite{schneider18}. While it is possible to fill the metering chamber multiple times and combine the fluid to create larger volumes, it is not possible to determine whether the chamber is filled to a certain degree, making it impossible to reliably meter volumes smaller than the size of the metering chamber. To assure correct functionality of the chip, all volumes used therefore have to be multiples of the HTR. In our examples, we will assume the HTR to be 1 nl.\\

\begin{figure}[!b]
    \centering
        \includegraphics[width=0.33\textwidth]{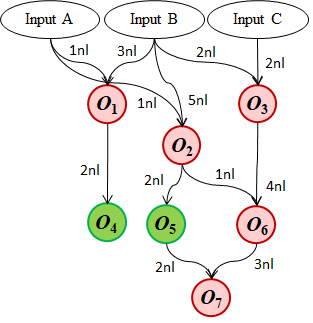}
        \caption{Application graph showing the input requirements of the operations as ratios for mixing operations (red) and as discrete values for the other type of operations (green).}
        \label{fig:applicationnonotes}
\end{figure}

\subsection{Application Model}
\noindent To model a biochemical application, we use a directed,
acyclic and polar sequencing graph, as explained in detail
in~\cite{pop2016microfluidic}. A node in the graph is a microfluidic operation that runs on a FFU. The edges denote dependencies between operations that require fluid transport.
An example of such an application graph can be seen in Fig.~\ref{fig:applicationnonotes}. Operations $O_1$ to $O_3$, $O_6$ and $O_7$ are mixing operations and their mixing ratio is indicated on the incoming edges, for examples $1/4$ and $3/4$ for $O_1$. $O_4$ and $O_5$ are detection operations and the label on the incoming edge denotes the MVR to successfully perform the detection. Operations may produce excess fluid (waste), which is not required by the immediate successors. This is the case whenever the sum of required input fluids is larger than the sum of fluid required by following operations.
Application models are not architecture specific. Therefore, before an application can be run on a biochip, the operations have to be mapped to the chip. This includes binding of operations, routing and scheduling~\cite{pop15}.

\section{Problem Formulation}

The inputs to our problem are (1) the biochip architecture model (including the MHC, the MVRs of FFUs and the HTR), (2) the biochemical application model, including the mixing ratios, and (3) the available technology (e.g. 1:1 or arbitrary ratio mixers). We are interested in determining the \textbf{Fluid Volume Assignment (FVA)} for each operation of the application such that all the fluidic constraints are satisfied and the total fluid volume is minimized. Additionally, the use of input fluids is to be reduced further by reusing waste generated by the operations.

\section{Mixing}
\label{sec:mixing}

For many biochemical applications the most common operation is mixing, be it in preparation of an experiment to produce a Master-Mix, preparation for detection requiring to add dye to the samples or long term experiments requiring additional reagents at various times. At the same time, mixing operations are the main source of fluid waste in continuous-flow biochips as mixing operations are prone to produce excess fluid. For efficient volume management while using such mixing hardware, we therefore first review mixing optimizations presented in related work.\\

Research so far has provided multiple solutions for optimizing mixing operations, both for droplet and flow based biochips~\cite{Dinh:Mix}~\cite{roy15}~\cite{Liu:Mix}. These techniques however are focused on optimizing mixing operations to require as few mixing steps as possible, which can increase fluid consumption and require high computational complexity. The waste-aware mixing algorithm proposed in~\cite{thies08} allows to minimize the consumption of one input fluid, but is restricted to a mix of only two fluids and drastically increases the consumption of the other fluid. Furthermore,~\cite{thies08} focuses on optimizing single mixing operations as opposed to the application-wide fluid management problem we address in this paper.

\begin{figure}[!t]
    \centering
        \includegraphics[width=0.48\textwidth]{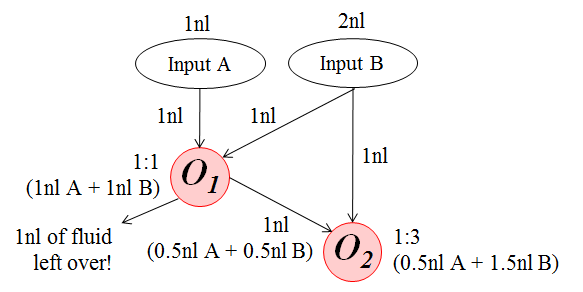}
        \caption{Example of achieving a 1:3 mixing ratio using 1:1 mixing hardware and the unavoidable byproduct of fluid (i.e. waste) not in the target ratio.}
        \label{fig:mixing}
\end{figure}

\begin{figure*}[!b]
    \centering
    \subfloat[Min-Mix]{
    \includegraphics[width=0.25\textwidth]{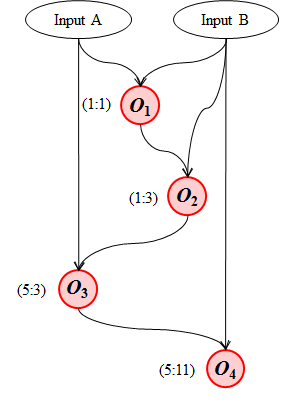}
    \label{fig:mixMM}
    }
    \subfloat[REMIA]{
     \includegraphics[width=0.25\textwidth]{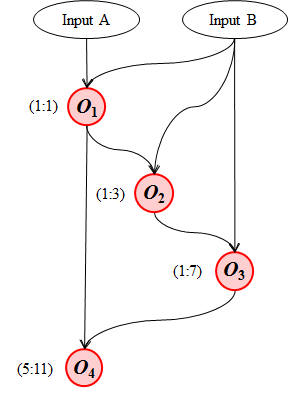}
    \label{fig:mixREMIA}
    }
    \subfloat[NFB]{
     \includegraphics[width=0.25\textwidth]{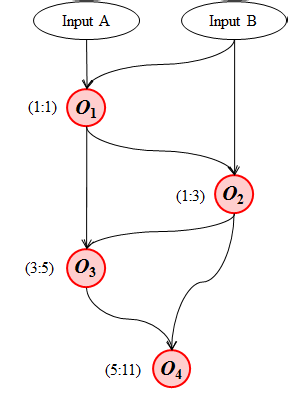}
    \label{fig:mixNFB}
    }
    \caption{Comparison of mixing trees for a target ratio of 5:11 created by Min-Mix, REMIA and NFB}
        \label{fig:mixingcomparison}
\end{figure*}

\noindent The 1:1 mixing architecture as shown in Fig.~\ref{fig:mixer} mixes two equal volumes of fluid. To achieve ratios other than 1:1, multiple sequential mixing steps are necessary. Fig.~\ref{fig:mixing} shows how a 1:3 mixing ratio can be achieved by adding a second mixing operation which mixes the result of the first operation with additional fluid from the input. However, the architecture of 1:1 mixers makes sequential mixing such as this especially wasteful. The output volume of a mixing operation is the sum of both input volumes, therefore if O$_1$ receives 1 nl from each input, it produces an output volume of 2 nl. Assuming the same size of mixer for all operations, only 1 nl can be passed on to O$_2$ to be mixed with more fluid from the input, leaving the other 1 nl to be discarded. Using mixers of increasing sizes is infeasible. While using a mixer that takes two inputs of 2 nl for O$_2$ would allow O$_1$ to pass on all its output volume, it would also require Input B to dispense 2 nl instead of 1 nl to this operation, increasing the overall fluid consumption and furthermore just passing on the problem to a potential next mixing operation which would then require an even larger mixer.

\begin{figure}[!t]
    \centering
    \subfloat[Network graph with precision level 4]{
    \includegraphics[width=0.48\textwidth]{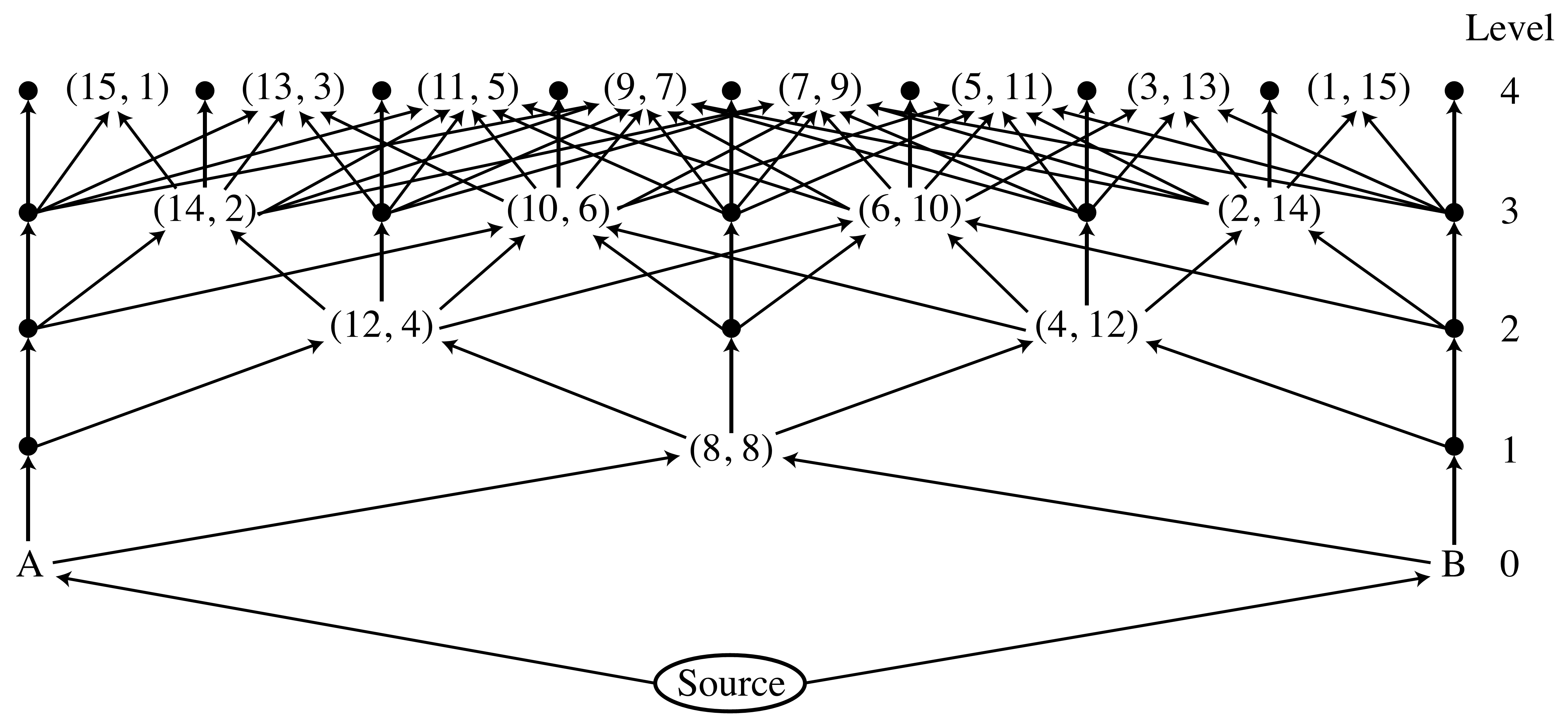}
    \label{fig:changeA}
    } \newline
    \subfloat[NFB solution for Fig.~\ref{fig:mixNFB}]{
     \includegraphics[width=0.48\textwidth]{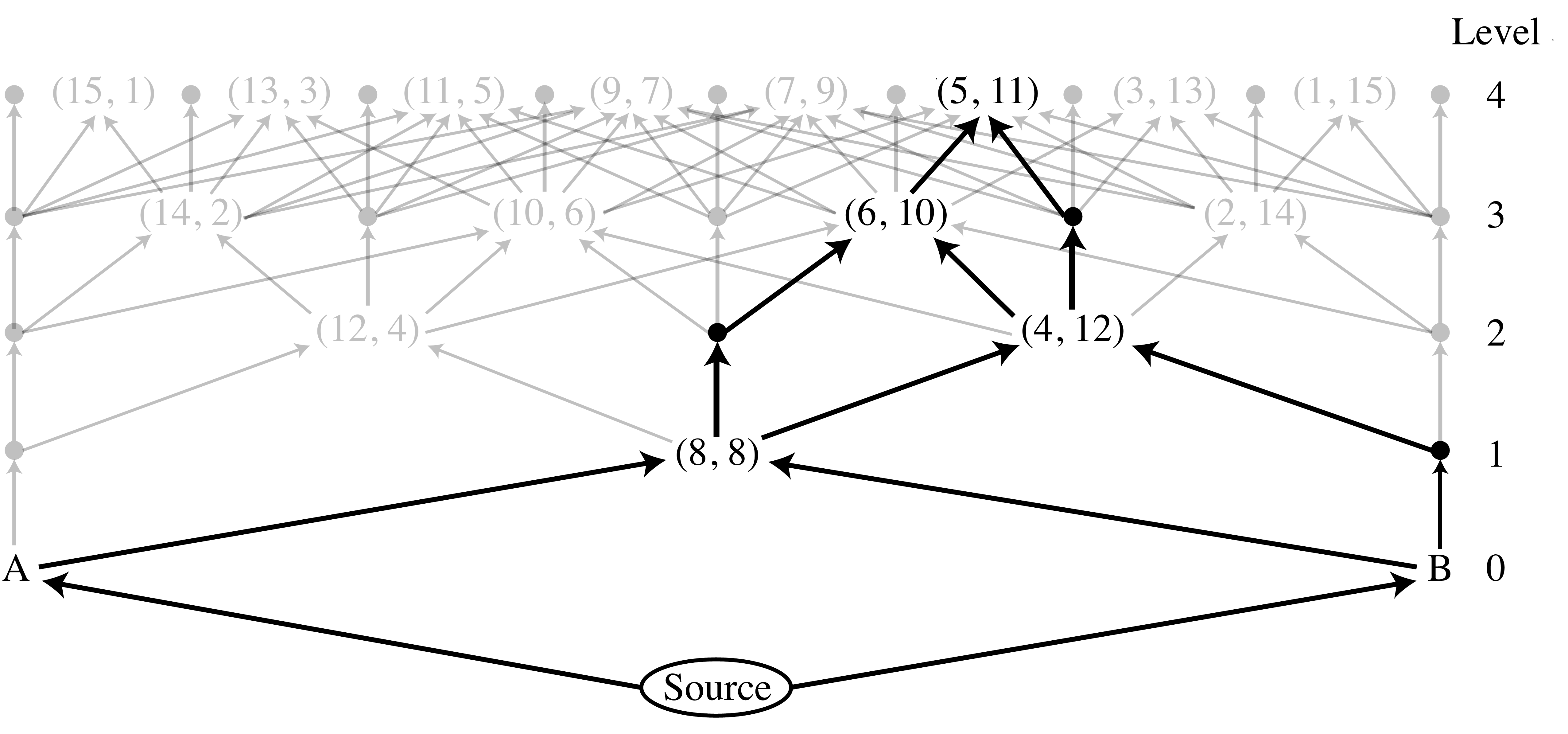}
    \label{fig:changeB}
    }
    \caption{Comparison of mixing trees for a target ratio of 5:11 created by Min-Mix, REMIA and NFB}
        \label{fig:NFBFull}
\end{figure}

Furthermore, not all mixing ratios are obtainable by chaining multiple 1:1 mixing operations together. Ratios which are unreachable using 1:1 mixing hardware are approximated and the quality of the approximation can be controlled as it is directly linked to the number of mixing operations performed, also called precision level~\cite{Dinh:Mix}. As mentioned before, sequential mixing is wasteful and should therefore be done as efficiently as possible. Multiple solutions have been proposed to find optimized mixing trees, such as Min-Mix (MM)~\cite{Thies:MM}, REMIA~\cite{Liu:Mix} and the Network Flow Based (NFB) algorithm~\cite{Dinh:Mix}. We have compared and evaluated each of the techniques and a basic example outcome for each can be seen in Fig.~\ref{fig:mixingcomparison} which also provides a direct comparison. While MM and NFB focus on minimizing the operation count, REMIA also considers the difference in cost of the inputs (e.g. sample and buffer) and minimizes the use of the expensive fluid. As Fig.~\ref{fig:mixingcomparison} already indicates, NFB performs superior regarding fluid consumption compared to MM and REMIA, as it solves an Integer Linear Programming (ILP) problem optimally, which can be visualized with network graphs such as shown in Fig.~\ref{fig:NFBFull}. Solving an ILP program does however require significantly more computation time than MM or REMIA. This can be tolerated to some extent, since the application graph optimization is generated during the design phase of the biochip, but large applications or very high approximation precision levels can still lead to infeasible calculation times.

\begin{figure}[!b]
    \centering
        \includegraphics[width=0.49\textwidth]{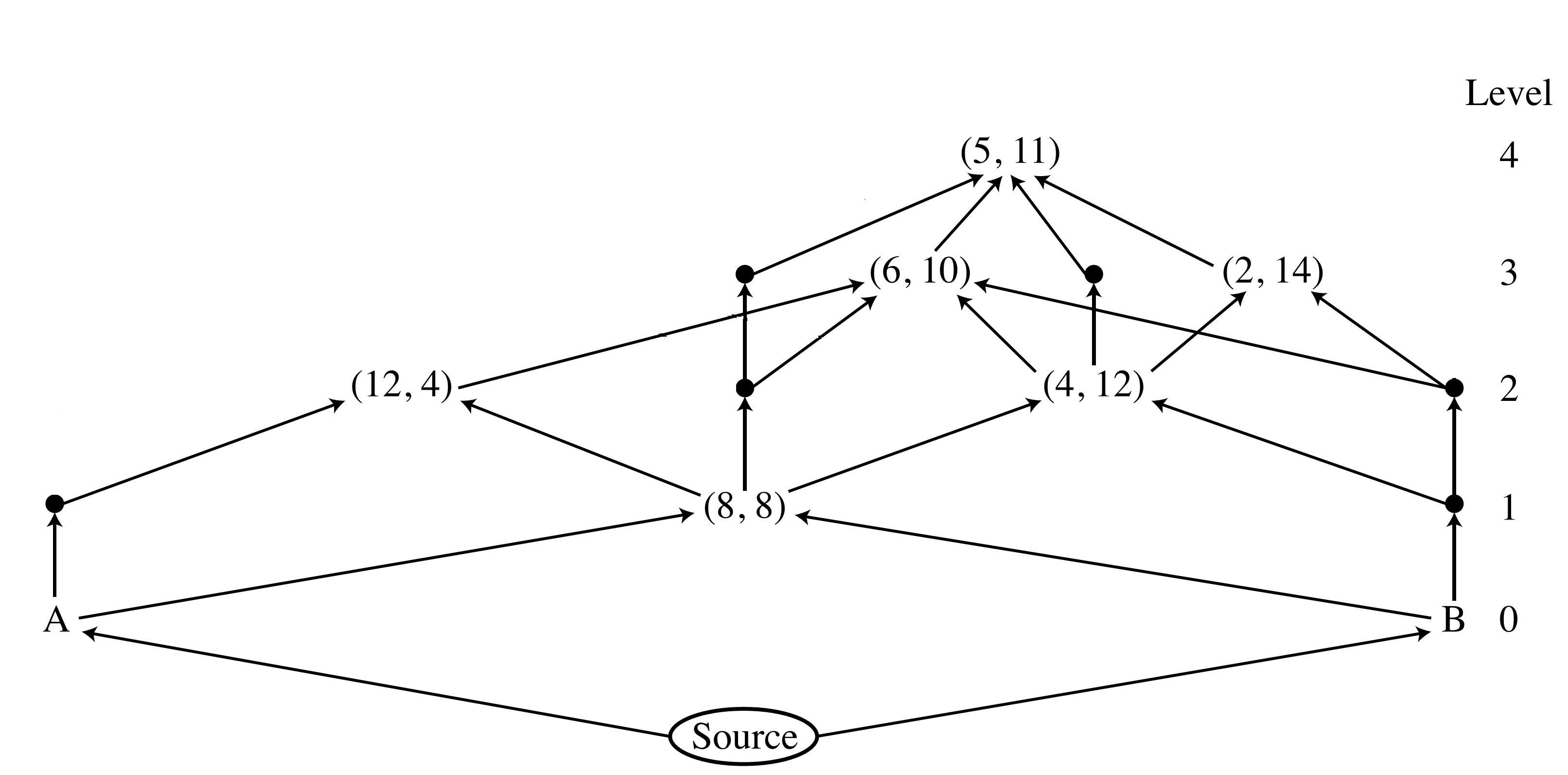}
        \caption{Application from Fig.~\ref{fig:applicationnonotes} with fluid volume according to each operations MVR assigned, leading to shortage of fluid at $O_6$.}
        \label{fig:NFBreduced}
\end{figure}

We therefore apply a pruning technique to the Network Graph, which removes a significant amount of vertices which are unlikely to be in the optimal solution. The resulting Network Graph is shown in Fig.~\ref{fig:NFBreduced}. The applied restriction allows no more than four vertices at any level $X$ to be used as input for any vertex at level $X + 1$. While this alteration no longer guarantees an optimal outcome, the results are still significantly superior to both MM and REMIA, and keeps the execution time feasible even for high precision levels as discussed in Sect.~\ref{sect:experiments}.

The computational complexity of NFB does therefore not pose an issue and we consider that the NFB4 (NFB pruned to 4 vertices) version of NFB provides the best solution to the mixing problem for our case and the mixing trees of our following examples have been created using this technique.

\begin{table*}[!t]
    \begin{center}
    \begin{tabular}{ |c|c|c|c|c|c|c|c|c|c| }
     \hline
      & \multicolumn{3}{|c|}{Precision Level 8} & \multicolumn{3}{|c|}{Precision Level 9} & \multicolumn{3}{|c|}{Precision Level 10}\\
      \hline
    Algorithm & MM & REMIA & NFB4 & MM & REMIA & NFB4 & MM & REMIA & NFB4 \\
      \hline
    Fluid Cost & 219.3 & 135.2 & 49.3 & 278.7 & 169.7 & 56.1 & 334.6 & 178.9 & 56.2 \\
      \hline
    Operations & 59.4 & 76.2 & 41.4 & 73.0 & 100.6 & 51.3 & 85.3 & 107.9 & 59.3 \\
      \hline
    Time (s) & 0.01 & 0.02 & 1.19 & 0.01 & 0.03 & 1.36 & 0.01 & 0.04 & 6.74 \\
     \hline
    \end{tabular}
    \end{center}
    \caption{Comparison of fluid consumption, number of operations and execution time for MM, REMIA and NFB4 (NFB pruned to 4 vertices), at precision levels 8, 9 and 10. All results represent the average outcome of 10 test cases, 8 synthetic and 2 real world applications}
    \label{tab:Mixing}
    \end{table*}

We have implemented and compared Min-Min (MM), REMIA, and the Network Flow Based (NFB) algorithms. The results are shown in Table~\ref{tab:Mixing}, which incorporates the data of multiple test cases at three different precision levels for all three optimization techniques. We determined that NFB, even if pruned to four vertices, and therefore no longer optimal but significantly faster, noticeably outperforms both MM and REMIA in both number of operations and fluid consumption.

An alternative to fixed mixing architectures are arbitrary ratio mixers as proposed in~\cite{Amin2013}~\cite{schneider18} and used in our previous volume management solution~\cite{schneider17}. Such mixers can drastically reduce the amount of excess fluid created during sequential mixing operations since most ratios can be mixed in a single operation. This is possible because the mixer does not require to be completely filled with fluid, but can be partially filled with air. Air which is picked up by the fluid during the mixing process can then be expelled through vents~\cite{Chuang:Vent}.
A ratio of 1:3 as shown in Fig.~\ref{fig:mixing} can be mixed in a single operation as any volume of fluid (up to the mixers maximum capacity) can be dispensed into the mixer. While this eliminates the wasteful in-between step which is necessary when using 1:1 mixers, the outcome is not necessarily more efficient. Due to the MHC as explained in Sect.~\ref{sect:ArchModel} larger volumes may have to be dispensed. In the example in Fig.~\ref{fig:mixing} a total of 1 nl Input A and 2 nl Input B is required to achieve the 1:3 mixing ratio and results in 2 nl of the target 1:3 ratio and 1 nl of byproduct in 1:1 ratio. Considering the MHC to be 1 nl, then using an arbitrary ratio mixer results in the mix of 1 nl of Input A and 3 nl of Input B. The overall fluid consumption is therefore higher, however 4 nl of the target 1:3 ratio is produced.
In addition to the prolonged exposure of the reagents to air, arbitrary ratio mixers come with other downsides. Longer execution times (due to slow venting) and complex fabrication make this technology more of an alternative than a clearly superior solution to fixed ratio mixers. We therefore provide examples of the volume management technique applied to architectures using fixed ratio mixers in the next section and compare the results to our previous data, gathered from architectures using arbitrary ratio mixers.

\section{Volume Management}
\label{Sect:VM}

\begin{figure}[h]
    \centering
        \includegraphics[width=0.45\textwidth]{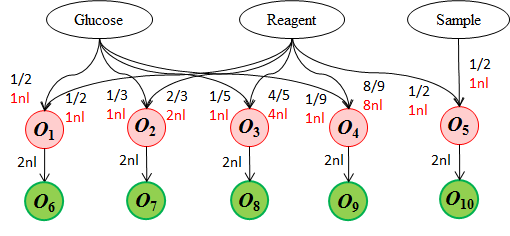}
        \caption{Application graph for a glucose test. Black labels indicate the required ratios and discrete volumes, red labels indicate the FVAs}
        \label{fig:glucose}
\end{figure}

\begin{figure}[h]
    \centering
        \includegraphics[width=0.48\textwidth]{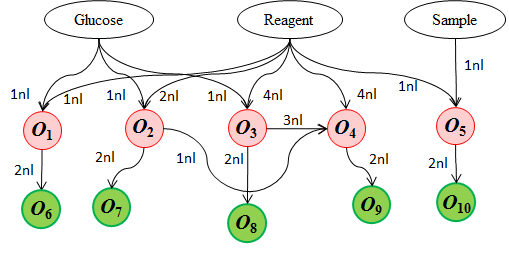}
        \caption{Optimized application graph, where left over fluid from O2 and O3 are used in O4. All values indicate FVAs.}
        \label{fig:glucosesolved}
\end{figure}

Automatic volume management allows to determine the required volumes of fluids according to the biochemical assay. Due to the technological and size differences, such volumes are likely to differ from those given in the assay. Instead, biochip related restrictions such as the HTR and MHC along with the mixing ratios given in the assay will determine the application's volume requirements as shown in Fig.~\ref{fig:applicationnonotes}. Discrete volumes for all inputs are then determined leading to volumes as shown in Fig.~\ref{fig:applicationnonotes}, which represent the smallest assignable volumes in accord with the given restrictions. This process is explained in detail in~\cite{schneider17}. In the same paper we describe how the use of fluids can be optimized, by reintroducing LeftOver Fluids (LOF) created by mixing operations to other operations within the application. As shown in a sample glucose application in Figs.~\ref{fig:glucose} and \ref{fig:glucosesolved}, a significantly smaller volume of fluids is required if leftover fluids are reused.

In this section we will present our automatic volume management algorithm in detail and provide examples using arbitrary mixing ratio architectures.  In our previous work, we have considered volume management for fixed output volumes and we provide solutions that satisfy all constraints while minimizing the total fluid volume required~\cite{schneider17}. However, this work is limited to architectures using arbitrary ratio mixing components.

\begin{figure}[!b]
    \centering
        \includegraphics[width=0.48\textwidth]{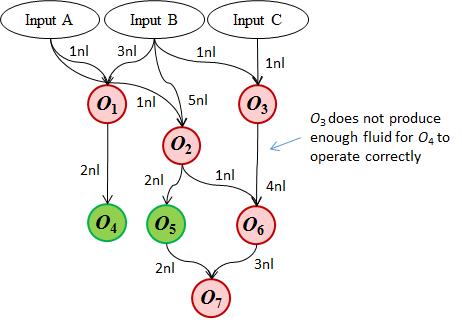}
        \caption{Application from Fig.~\ref{fig:applicationnonotes} with fluid volume according to each operations MVR assigned, leading to shortage of fluid at $O_6$.}
        \label{fig:applicationMVR}
\end{figure}

Let us consider the application from Fig.~\ref{fig:applicationnonotes} to be executed on the biochip from Fig.~\ref{fig:arch}. We assume a MHC of 10 nl for all FFUs and a HTR of 1 nl. We propose a method that calculates the minimal FVA for every operation and then optimize the input fluid consumption by reusing waste. This method is implemented in Alg.~\ref{algo}. Applying this algorithm to the application in Fig.~\ref{fig:applicationnonotes} results in FVAs as shown in Fig.~\ref{fig:applicationMVR}. In lines 1-12 the algorithm determines the minimal FVA for every operation in the application in reverse topological order.

To determine the minimal FVAs, every operations MVR has to be known. A MVR represents a local minimal volume assignment as it is the smallest amount of volume required to successfully execute a single operation. While MVRs of FFUs such as detectors are already discrete amounts that can be dispensed, for mixing operations the MVRs that obey the hardware restrictions have to be determined from the given mixing ratios first. We calculate the MVR for mixing operations as follows: The smallest amount of fluid that satisfies both the HTR as well as the target mixing ratio is the sum of the products of the HTR and the ratio numerators, e.g. for $O_1$ from Fig.~\ref{fig:applicationnonotes} which requires $1/4$ Input A and $3/4$ Input B we calculate $HTR * NumeratorA + HTR * NumeratorB = 1nl * 1A + 1nl * 3B$ resulting in an MVR of 1 nl of Input A and 3 nl of Input B.

\begin{algorithm}
\caption{Fluid Volume Assignment}
Input: Application graph, HTR, MHC, MVR for each FFU
\label{algo}
\begin{algorithmic}[1]
\FOR{each node $n$ of the application in rev. order }
\FOR{each outgoing edge $e$ of $n$ }
\STATE RF += $e$.RequiredFluid
\ENDFOR
\STATE FVAs = n.MVRs
\FOR{X = 1; FVAs $<$ RF; X++}
\STATE FVAs = MVRs $\diamond$ (HTR * X)
\ENDFOR
\IF{FVA $>$ MHC}
\STATE Cascading or Static Replication
\ENDIF
\ENDFOR
\FOR{each node $n$ of the application}
\IF{$n.CombinedInput > n.CombinedOutput$}
\STATE $LeftOverFluids.Add(n.LeftOverFluids)$
\ENDIF
\ENDFOR
\FOR{each leftover fluid $LOF$ in $LeftOverFluids$}
\IF{At least one input of another operation can be completely or partially replaced by the $LOF$}
\STATE remove $LOF$ and update volume assignments and application graph
\ENDIF
\ENDFOR
\end{algorithmic}
Output: Optimized application graph
\end{algorithm}

The MVR does however not consider how much fluid other operations in the application will require to successfully execute. Fig.~\ref{fig:applicationMVR} shows the application from Fig.~\ref{fig:applicationnonotes} with MVRs assigned to each operation. This however leads to underflow at $O_6$ since this operation requires 4 nl of fluid from $O_3$, but $O_3$ only has a combined input, and therefore maximum output, of 2 nl. To prevent such scenarios, Alg.~\ref{algo} first determines, for every operation, how much fluid they have to pass on to following operations in lines 2-4. The algorithm does so in reverse topological order, making a single pass over all operations sufficient. Leafs of the application, which are calculated first, never have to pass on any fluid. Therefore they receive FVAs which are identical to their MVRs. For example $O_4$ in Fig.~\ref{fig:applicationnonotes} receives 2 nl from $O_1$ and similarly $O_7$ receives 2 nl from $O_5$ and 3 nl from $O_6$ which is the MVR calculated from the mixing ratios and shown in Fig.~\ref{fig:applicationMVR}.

\begin{figure}[h]
    \centering
        \includegraphics[width=0.48\textwidth]{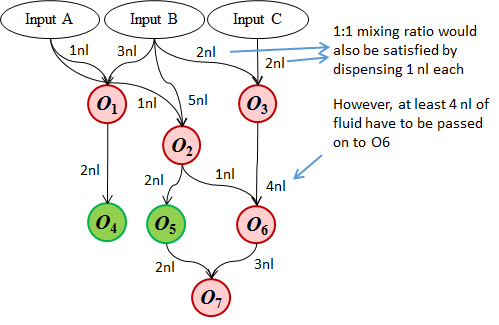}
        \caption{Application from Fig.~\ref{fig:applicationnonotes} with optimal assignment of fluid according to the calculated FVAs.}
        \label{fig:applicationgood}
\end{figure}

\begin{figure}[h]
    \centering
        \includegraphics[width=0.40\textwidth]{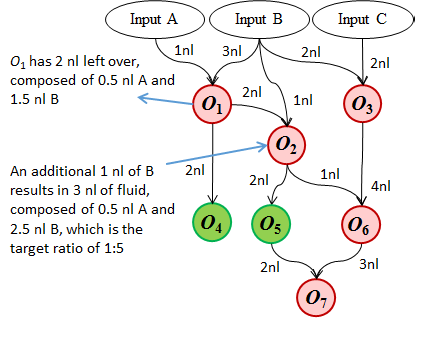}
        \caption{Application from Fig.~\ref{fig:applicationnonotes} with optimal assignment of fluid according to the calculated FVAs and leftover fluid from $O_1$ has been reused as input for $O_2$.}
        \label{fig:applicationLOF}
\end{figure}

All other operations have to assure that they process enough fluid to satisfy the following operations FVA which is checked in line 6-8. $O_6$ receives a total of 5 nl as input using its MVRs which is enough to satisfy the input of $O_7$. $O_3$ however would only produce 2 nl of fluid using its MVRs of 1 nl from Input B and 1 nl from Input C as shown in Fig.~\ref{fig:applicationMVR}. The algorithm then scales these inputs in line 7. We solve for $x$ in the following equation: $RequiredOutput <= \sum_{n=1}^{numMVRs} MVR_n \diamond (HTR * x), $for $ x \in \mathcal{N}$ where $\diamond$ stands for multiplication for mixing operations, which scales the input while keeping the mixing ratio, and addition for other operations which determines the smallest multiple of HTR that satisfies the required output. For our example of $O_3$ we can see in Fig.~\ref{fig:applicationgood} that both inputs have been scaled by a factor of two, keeping the mixing ratio as well as providing enough fluid to pass on to $O_6$.
After determining the FVA the algorithms assures that no overflow will occur in lines 9-12. Overflow occurs whenever more fluid than the MHC has to be dispensed i.e. the FVA is larger than the MHC. This issues can be resolved through cascading or static replication as described in detail in~\cite{amin08}.

Wasted fluids are to be reused whenever possible to further decrease the fluid consumption. Furthermore, we are interested to design a low computational complexity algorithm, to allow for non-deterministic events such as conditional execution or errors to be solved during the execution of the biochemical application.

As mentioned previously, 1:1 mixing hardware~\cite{pop15} has an inherit issue of producing waste for all ratios except for 1:1, as illustrated in Fig.~\ref{fig:mixing}. LOFs can however be assigned to any operation that is not a predecessor of the operation creating the LOF and that contains the same basic reagents. When solving applications such as the example in Fig.~\ref{fig:glucose} with 1:1 mixing hardware, not just the indicated mixing operations, but also all intermediate operations that are required to reach the target ratio with 1:1 mixers have to be considered.
To successfully reassign as much LOFs as possible, a thorough analysis of all given data is necessary as we will show using the example from Fig.~\ref{fig:glucose}. This example consists of an application graph of a glucose test where black labels indicate the MVR and the red labels the FVA. Here, $O_2$, $O_3$ and $O_4$ produce LOFs.

Our algorithm determines these LOFs in lines 13-17. The data required for optimal LOF reassignment consists of the LOF volume, the operation that creates it and the fluid composition regarding the applications original inputs, i.e. for the glucose example, how much Glucose, Reagent and Sample the LOF contains. $O_3$ from Fig.~\ref{fig:glucose} for example, receiving a total of 5 nl as input and passing 2 nl on to $O_8$, creates a LOF with a volume of 3 nl, containing 1.5 nl Glucose and 1.5 nl Reagent.

As previously mentioned, a LOF can be used whenever it allows to partially or completely remove one or more incoming edges of an operation, if the resulting volume still obeys the target operations FVA and the ratios of the contained fluids remains correct. However, it is also possible to only use part of the LOF, or combine multiple LOFs in order to meet these requirements. The assignment of LOFs containing multiple fluids is made possible by a large extend due to acceptable error margins when mixing fluids~\cite{alistarError}. One of these cases is shown for our example from Fig.~\ref{fig:glucose}. Using the initially calculated FVAs, operations $O_2$, $O_3$ and $O_4$ produce LOFs. The algorithm iterated through all possibilities to reassign these LOFs and determines that the LOF produced from operations $O_2$ and $O_3$ can be used as partial input for $O_4$, as shown in Fig.~\ref{fig:glucosesolved}. Here, $O_4$ no longer receives 1 nl of Glucose (G) and 8 nl of Reagent (R) from inputs, but instead 0.33 nl G and 0.66 nl R from $O_2$, 0.6 nl G and 2.4 nl R from $O_3$ and 4 nl R from an input. This corresponds to a total of 0.93 nl G and 7.06 nl R, or a ratio of 0.93:7.06 which is 0.6\% off the desired ratio of 1:8, which is within acceptable error margins. As a result O$_4$ consumes 1 nl G and 4 nl R less from inputs, reducing the overall volume requirements of the application as well as the need to route LOFs from $O_2$ and $O_3$ to waste ports.

\begin{figure}[t]
    \centering
    \subfloat[Mixing of ratio 3:13 creating excess fluid\newline of ratios 1:3 and 3:5]{
    \includegraphics[width=0.38\textwidth]{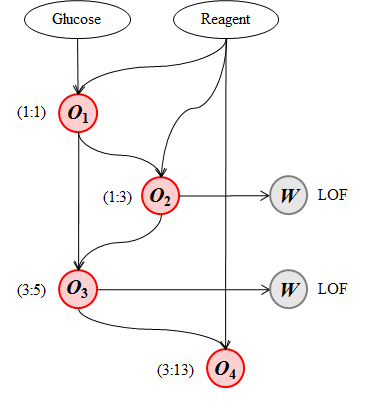}
    \label{fig:LOFdiff1}
    } \newline
    \subfloat[Mixing of ratio 3:13 creating excess fluid\newline of ratios 1:1 and 1:7]{
     \includegraphics[width=0.38\textwidth]{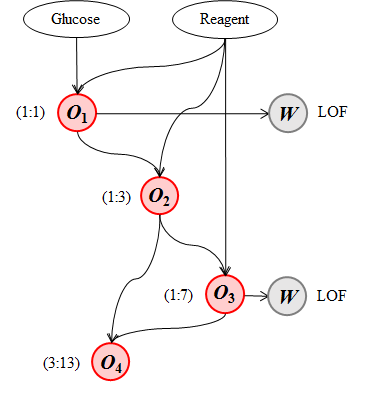}
    \label{fig:LOFdiff2}
    }
    \caption{Two mixing trees using the same input fluids to produce the same output fluids, but creating different LOFs due to different mixing order.}
    \label{fig:LOFdiff}
\end{figure}

Assuming 1:1 mixing hardware is used instead of arbitrary mixers, some extra steps are necessary for the fluid consumption to be optimized. We determine mixing trees as explained in Sect. \ref{sec:mixing} to obtain our target ratios. NFB provides a solution that requires the smallest volume of input fluids. However, multiple solutions consuming the same amount of fluid might be available. These solutions differ in the ratio of the LOF that they create during the production of the target ratio. Fig~\ref{fig:LOFdiff} shows an example of this where the overall fluid consumption and result are the same, but different ratios of the input fluids are left over. We therefore have NFB calculate all solutions that require the same amount of input fluid allowing our volume management algorithm to choose the mixing tree that produces the most useful LOFs.

\begin{figure}[h!]
    \centering
    \subfloat[Mixing tree created by NFB for ratio 1:1]{
    \includegraphics[width=0.38\textwidth]{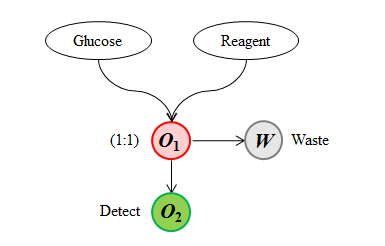}
    \label{fig:comparison11org}
    } \newline
    \subfloat[Optimized mixing tree for ratio 1:1 storing the LOF of O$_1$ for use in another mixing tree]{
     \includegraphics[width=0.38\textwidth]{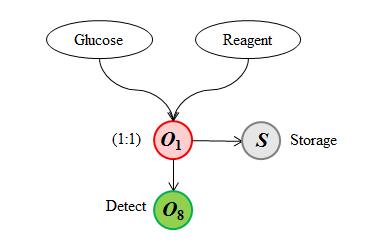}
    \label{fig:comparison11opt}
    }
    \label{fig:comparison11}
    \caption{Mixing trees for target ratio 1:1 which can be precisely mixed and no approximation is necessary}
\end{figure}

\begin{figure}[h!]
    \centering
    \subfloat[Mixing tree created by NFB for ratio 5:11]{
    \includegraphics[width=0.38\textwidth]{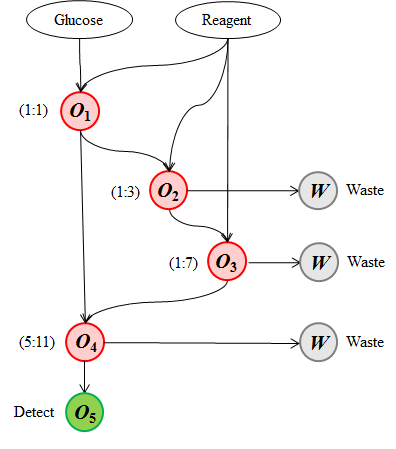}
    \label{fig:comparison12org}
    } \newline
    \subfloat[Optimized mixing tree for ratio 5:11 using LOF from O$_1$ and O$_6$ to reduce the use of G and R]{
     \includegraphics[width=0.38\textwidth]{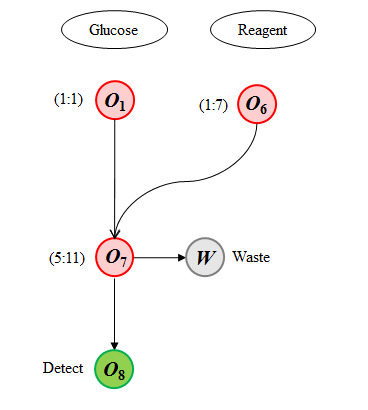}
    \label{fig:comparison12opt}
    }
    \label{fig:comparison12}
    \caption{Mixing trees for target ratio 1:2 which is approximated using a maximum of 4 operations to ratio 5:11}
\end{figure}

\begin{figure}[h!]
    \centering
    \subfloat[Mixing tree created by NFB for ratio 3:13]{
    \includegraphics[width=0.38\textwidth]{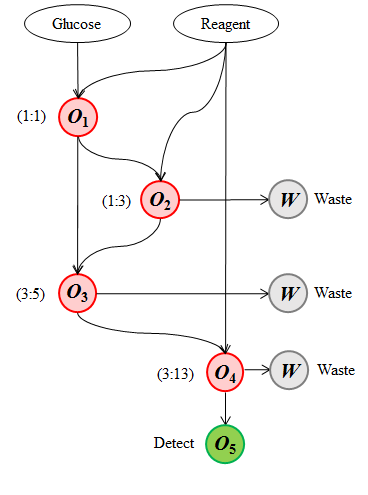}
    \label{fig:comparison14org}
    }
    \newline
    \subfloat[Optimized mixing tree for ratio 3:13 storing the LOF of O$_3$ for use in another mixing tree]{
     \includegraphics[width=0.38\textwidth]{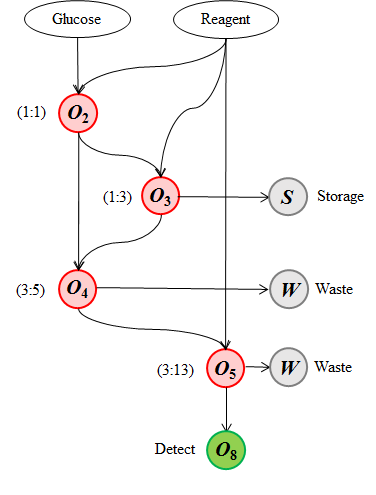}
    \label{fig:comparison14opt}
    }
    \label{fig:comparison14}
    \caption{Mixing trees for target ratio 1:4 which is approximated using a maximum of 4 operations to ratio 3:13}
\end{figure}

\begin{figure}[h!]
    \centering
    \subfloat[Mixing tree created by NFB for ratio 1:7]{
    \includegraphics[width=0.38\textwidth]{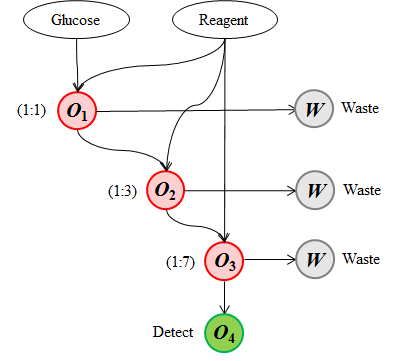}
    \label{fig:comparison18org}
    } \newline
    \subfloat[Optimized mixing tree for ratio 1:7 using LOF from O$_3$ to reduce the use of G and R and storing the LOF of O$_6$ for use in another mixing tree]{
     \includegraphics[width=0.38\textwidth]{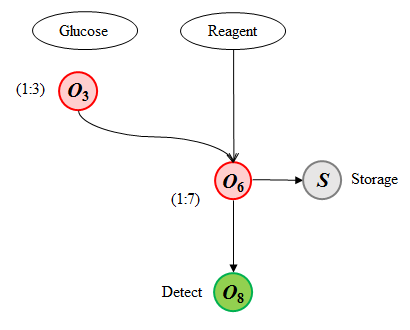}
    \label{fig:comparison18opt}
    }
    \label{fig:comparison18}
    \caption{Mixing trees for target ratio 1:8 which is approximated using a maximum of 4 operations to ratio 1:7}
\end{figure}

We apply this technique to our previous example from Fig.\ref{fig:glucose}, but only focus on the four required mixing ratios of G and R. As mentioned previously, 1:1 mixing hardware has the inherent problem that not all mixing ratios are obtainable. Some ratios therefore have to be approximated. The resulting error is dependent on the number of consecutive mixing operations, where more mixing operations allow for a smaller error in the approximated ratio. For our example, we choose a maximum number of mixing operations, or precision level, of 4 for each approximation. The resulting mixing trees are shown in Figs.~\ref{fig:comparison11org},~\ref{fig:comparison12org},~\ref{fig:comparison14org} and \ref{fig:comparison18org}. The target ratio 1:1 can be produced correctly, while the ratio 1:2 is approximated to 5:11, ratio 1:4 is approximated to 3:13 and ratio 1:8 is approximated to 1:7. These mixing trees are independent from each other and produce the target ratio by mixing reagents from the G and R input reservoirs. In most cases, the mixers output will not exactly match the detectors requirement and some fluid is discarded, e.g. a mixer with a capacity of 4 nl will pass on 2 nl to the detector and discard 2 nl.
The algorithm now minimizes the fluid consumption by reassigning the discarded fluids. The resulting mixing trees are shown in Figs.~\ref{fig:comparison11opt},~\ref{fig:comparison12opt},~\ref{fig:comparison14opt} and \ref{fig:comparison18opt}.
The mixing trees are no longer independent and the operation sequence is now shared across all mixing trees. However, using this setup and reassigning LOFs, the combined consumption of all four mixing trees is only 2 units of G and 6 units of R. In comparison, supplying the results of NFB with the required inputs without reassigning LOFs requires 4 units of G and 9 units of R.

\begin{table*}
\begin{center}
\begin{tabular}{ |c|c|c|c|c|c|c| }
 \hline
Assay & Minimal & NFB M$_1$ & NFB M$_2$ & Opt M$_1$ & Opt M$_2$ & Opt A \\
  \hline
CCA & 60 R; 190 B & 125 R; 300 B & 250 R; 600 B & 75 R; 225 B & 100 R; 300 B & 60 R; 200 B \\
  \hline
PAA & 30 R; 470 B & 250 R; 1050 B & 500 R; 2100 B & 100 R; 650 B & 100 R; 900 B & 30 R; 500 B\\
  \hline
\end{tabular}
\end{center}
\caption{Experimental results for optimization of a Colorimetric Cholesterol Assay (CCA) and a Proteasome Activity Assay (PAA). The table shows: 1) The assay name. 2) Minimal requirements of Reagent (R) and Buffer (B) to satisfy the target ratios and output volumes. 3) Consumption of R and B after approximation using NFB4 and assuming mixing hardware size that fits the required output volumes (i.e. 50 $\mu$l for CCA and 100 $\mu$l for PAA). 4) Same as 3, but assuming mixing hardware that fits double the volume. 5) Consumption of R and B after applying our FVA, assuming mixing hardware size that fits the required output volumes. 6) Same as 5, but assuming mixing hardware that fits double the volume.}
\label{tab:results}
\end{table*}

\section{Experimental Evaluation and Discussion}
\label{sect:experiments}

We have shown in our examples in Sect.~\ref{Sect:VM} and from related work that the volume of waste created by both arbitrary and 1:1 mixing hardware can be reduced significantly. We determined that the NFB algorithm provides superior solutions to MM and REMIA which we will verify in this section with experimental results. Furthermore will we compare the fluid consumption for these experiments using architecture with either fixed, or arbitrary ratio mixers. \\

In the next set of experiments, we have evaluated the Fluid Volume Assignment (FVA) algorithm from~\cite{schneider17} on real-life assays to validate the improvements of mixing tree generation using NFB.
We present the results of two additional assays, namely a Colorimetric Cholesterol Assay (CAA) as described in~\cite{cholesterol} and a Proteasome Activity Assay (PAA) as described in~\cite{proteasome}. The experimental results for these assays can be seen in Table~\ref{tab:results}. For simplicity, we refer to the fluids used in both assays as Reagent (R) and Buffer (B), the actual components used in these assays can be found in the respective assay protocols. Both assays require 5 samples of R and B mixed at different ratios. For CAA the desired output volume for each sample is 50 $\mu$l and for PAA 100 $\mu$l.
The CCA target ratios have been approximated to a precision level of 4, resulting in the approximated ratios: 1:3, 1:15, 3:13, 5:11 and 3:5 of R:B respectively.
The PAA target ratios have been approximated to a precision level of 6, resulting in the approximated ratios: 1:7, 3:29, 5:59, 1:15 and 1:31 of R:B respectively.
As Table~\ref{tab:results} shows, the FVA technique can reduce the required volumes significantly, for CAA the use of R and B are reduced to 60\% and 75\% and for PAA to 40\% and 62\%, respectively, compared to the results obtained using the mixing trees obtained from NFB without optimization. These experiments assume mixing hardware which has a capacity that exactly matches the desired output volume for each assay, marked M$_1$ in the table (i.e. 50 $\mu$l for CCA and 100 $\mu$l for PAA).\\ We also determined the results for the same assays for the case when larger capacity mixing hardware is available, with a volume double to the desired output volume of the assays (i.e. 100 $\mu$l for CCA and 200 $\mu$l for PAA). While the absolute fluid consumption rises as is to be expected when using larger mixing hardware, the optimization factor increases as well compared to the previous results. Using these larger mixers (the results marked M$_2$ in Table~\ref{tab:results}) we obtain a reduction of R and B in CAA to 40\% and 50\% and for PAA to 20\% and 43\% respectively, compared to unoptimized NFB results, using such larger mixers as well. This increased optimization is an advantage if architectures are used which are not specifically designed for one assay, but are general purpose and might therefore use larger mixers. \\

While the required fluid volumes compared to the naive NFB solutions are significantly smaller, there is still a noticeable gap to the minimal requirements. The M$_1$ solution of CCA requires 25\% more R than the minimum and the M$_1$ solution of PAA requires more than 300\%. This is essentially the same Hardware Transport Resolution (HTR) issue as discussed before. Due to the mixers fixed chamber size, only multiples of certain volumes can be used. For example, CAA requires an output volume of 50 $\mu$l. The used mixers therefore have a capacity of 50 $\mu$l total, or 25 $\mu$l per mixing chamber, which can be considered the HTR for this case. With 75 $\mu$l being the smallest multiple in the HTR that is larger than the minimal requirement of 60 $\mu$l, the optimization is only hindered by the hardware restriction. The PAA optimization suffers from the same issue, however as it occurs multiple times throughout its mixing tree, the fluid consumption is significantly higher than the minimal fluid requirements. To improve these results, additional mixing hardware would be required with different capacities (e.g. one mixer of 50 $\mu$l and one of 25 $\mu$l capacity) and/or different ratios (e.g. one mixer with 1:1 ratio and one with 1:2 ratio). This would however negatively impact the experiment in several ways, such as number of operations, architecture complexity and execution time.\\

As shown in~\cite{schneider17}, arbitrary ratio mixers only produce waste resulting from skewed mixing ratios and high HTRs. Therefore, minimal consumption solutions for our examples can be obtained using such hardware. However, for CAA for example, this would require a small HTR of 2 $\mu$l, resulting in the need for smaller metering components and additional metering steps. Our architecture using fixed ratio mixers in contrast can use a HTR of 4 $\mu$l. Therefore, to obtain a fair comparison, the target ratio using arbitrary ratio mixers can be approximated to a certain extend, similarly to the solution for fixed ratio mixers to allow for a HTR of 4 $\mu$l as well. Comparing the use of buffer for the minimal consumption and the arbitrary mixing ratio solution in Table.~\ref{tab:resultsarb}, it is notable that the approximation is achieved by increasing the amount of buffer used. The volume of reagent remains the same as all volumes are already multiples of 4. However, even if this is not the case, the additionally required volume for any reagent for arbitrary ratio mixer approximation is significantly smaller compared to fixed ratio mixer requirements as illustrated by the small increase in buffer requirements in our examples.

\begin{table}
\begin{center}
\begin{tabular}{ |c|c|c|c| }
 \hline
Assay & Minimal & Fixed & Arbitrary \\
  \hline
CCA & 60 R; 190 B & 75 R; 225 B & 60 R; 200 B \\
  \hline
PAA & 30 R; 470 B & 100 R; 650 B & 30 R; 500 B \\
  \hline
\end{tabular}
\end{center}
\caption{Comparison of fluid consumption for fixed and arbitrary ratio mixer hardware regarding the CAA and PAA assays}
\label{tab:resultsarb}
\end{table}

\section{Related Work}
\label{Sect:related-work}

Recent advancements in fluid-volume assignment for continuous-flow microfluidic biochips (CFMBs) have focused on integrating realistic fluid-manipulation constraints into the physical design phase, developing contamination-resilient switch architectures, and algorithmically optimizing reagent dilution networks. A review by Huang et al.~\cite{huang2021computer} highlights the evolution of computer-aided design techniques for flow-based microfluidic systems, emphasizing the growing importance of volume management in modern biochip design. However, ensuring precise volume metering and transport under laminar-flow regimes imposes ongoing challenges related to dispersion, device integration, and control-layer complexity.

Determining the required fluid volumes for simple biochemical assays can be straightforward and due to the inherently small volumes required for biochips, wasted fluid might be considered acceptable for certain reagents. However, for a growing number of applications, dispensing more fluid than needed (overflow) can be expensive, while shortages of fluid (underflow) can interrupt the execution and even require manual replenishment. Furthermore, hardware restrictions can impose constraints on the amount of fluid that can be passed on from one operation to another, details of which are presented in Sect.~\ref{sect:ArchModel}. Additionally, non-deterministic events such as errors~\cite{alistarError} or conditional execution~\cite{roy15} can further complicate fluid management.

Traditional CFMB synthesis approaches separate scheduling, binding, and physical-design steps, often neglecting the interplay between routing paths and fluid-volume requirements. Liu et al. \cite{liu2024three} introduced a three-stage rapid physical-design algorithm that explicitly models fluid-volume consumption during routing, storage, and mixing operations. In the first stage, fluid-handling tasks are clustered by volume demand and timing to minimize intermediate storage; the second stage allocates channel segments proportional to these volumes while respecting valve-actuation sequences; the final stage performs layout compaction without violating fluid-volume capacities on each channel segment. By co-optimizing volume assignment and channel-length minimization, their method achieves up to 30\% reduction in chip area and 25\% fewer valve actuations compared to legacy approaches.

Previous research on automatic volume management has proposed a linear programming solution, which is capable of avoiding over- and underflow for systems with arbitrary mixing ratio hardware. However, the execution time is infeasible for large assays, hence a fast heuristic is proposed for an over-constrained version of the problem, which in turn does not guarantee to provide a valid solution~\cite{amin08}. Furthermore, the authors are interested in a solution that maximizes the sum of the output volumes and ignore the waste produced during the execution.

Cross-contamination during reagent routing remains a primary concern when reusing channels for fluids of differing compositions. Shen et al. \cite{shen2022contamination} proposed a contamination-free switch design that integrates dedicated priming channels and switch-state encoding to prevent residual fluid volumes from entering subsequent flow paths. Each switch employs isolation buffers sized to accommodate the dead-volume of upstream channels, effectively decoupling fluid-volume assignment from switch reuse patterns. Their synthesis framework ensures that volume quotas reserved for buffer flushing are calculated during control-layer synthesis, leading to zero-contamination operation under worst-case volume scenarios.

Precise on-chip dilution series demand exact fluid-volume metering across multiple inlet channels. Banerjee et al. \cite{banerjee2021sample} exploited Farey sequencing to generate mixing networks that realize arbitrary dilution ratios using a minimal set of fixed-volume modules. By decomposing a target ratio into a sequence of integer partitions, their design synthesizes a network of constant-volume splits and merges, ensuring that each module handles a predetermined fluid volume. This technique drastically reduces the combinatorial complexity of volume-assignment decisions, enabling automated synthesis of sample-preparation modules with guaranteed volume-accuracy within 2\% of the target.

For applications demanding sample volumes beyond the microliter range, Etxebarria-Elezgarai et al. \cite{etxebarria2020large} demonstrated large-volume, self-powered disposable cartridges by integrating modular polymer micropumps that autonomously metered up to 500 µL of fluid through capillary-driven channels. Their design assigns fluid-volume segments by spatially partitioning the network into equal-pressure zones, each calibrated to a specific volume quota. This approach circumvents external pressure sources and simplifies volume assignment by standardizing channel dimensions and pump-actuation profiles.

Maintaining precise volume assignments during high-speed continuous flow requires accounting for axial dispersion phenomena. Motahari et al. \cite{motahari2024continuous} developed an inlaid microfluidic platform with solenoid-valve injection that dynamically adjusts injection volumes based on real-time Taylor–Aris dispersion modeling. They simulated dispersion effects for various injection volumes at constant flow rates and validated that volume assignment errors remained below 1\% when compensating for dispersion-induced broadening. Such feedback-augmented volume control represents a significant step toward error-resilient fluid assignment in CFMBs.

\section{Conclusion}
\label{sect:conclusion}

This paper examined fluid volume assignment methods for flow-based biochips, comparing different mixing approaches and their effects on fluid use. Our work shows that proper volume management prevents fluid shortages and excess while reducing waste through leftover fluid reuse. We tested several mixing tree algorithms and found that the modified NFB method (NFB4) offers the best balance between solution quality and calculation speed. Our tests with real biochemical assays showed that our approach cuts fluid use by 40-80\% compared to basic methods. While arbitrary ratio mixers use less fluid than fixed ratio (1:1) mixers, both benefit from our optimization techniques. The methods presented here work with various microvalve technologies and help make biochips more practical by reducing reagent costs and minimizing waste.

\balance
\bibliographystyle{IEEEtran}
\bibliography{refs}

\end{document}